\documentclass[]{aa}
\usepackage{graphicx}
\usepackage{multirow}
\usepackage{txfonts}
\titlerunning{Spectroscopy and polarimetry of Q0957+561}
\authorrunning{L. \v C. Popovi\'c, et al.}

\begin{document}

   \title{Spectroscopy and polarimetry of  the gravitationally  lensed quasar Q0957+561}


   \author{L. \v C. Popovi\'c
   \inst{1,2} 
   \and
   V. L. Afanasiev
   \inst{3}
   \and
   E. S. Shablovinskaya
   \inst{3}
   \and
   V. I. Ardilanov
   \inst{3}
   \and Dj. Savi\'c \inst{1}
  }

    \institute{Astronomical Observatory, Volgina 7, 11000 Belgrade, Serbia \\
    \email{lpopovic@aob.rs}
    \and
    Department of astronomy, Faculty of mathematics, University of Belgrade Studentski trg 16, Belgrade, 11000, Serbia\\
\email{lpopovic@matf.bg.ac.rs}
   \and
    Special Astrophysical Observatory of the Russian AS, Nizhnij Arkhyz, Karachaevo-Cherkesia 369167, Russia\\
   \email{vafan@sao.ru,gaerlind09@gmail.com} }

   \date{Received November 15, 2020; accepted January 18, 2021}


  \abstract
   {We present new spectroscopic and polarimetric observations of  the first discovered gravitational lens Q0957+561. The lensed quasar has been observed with the 6m telescope of the Special Astrophysical Observatory (SAO, Russia) in polarimetric and spectroscopic modes.}
   {We explore spectropolarimetric parameters of Q0957+561 A,B components to investigate the innermost structure of gravitationally lensed quasar, and explore the nature of polarization in lensed quasars. Additionally, we aim to compare their present-day spectral characteristics  with previous observations in order to study long-term spectral changes.}
   {We perform new spectral and polarization observations of Q0957+561 A,B images. After observed data reduction, we analyze spectral characteristics of lensed quasar comparing spectra of  A and B images, as well as comparing previously observed image spectra with present-day ones. The polarization parameters of the two images are compared. We also model the macro-lens influence on the polarization of the images representing the gravitational lens with a singular isothermal elliptical potential.}
   {We find that the brightness and SED ratio of components A and B  changed during a long period. Polarization in broad lines of components A and B showed that the equatorial scattering {   cannot be detected} in this lensed quasar.  We find wavelength-dependent polarization that may be explained as a combination of the polarization from the disc and outflowing material.
   There is a significant difference between polarization parameters of the A and B images, where the B component shows a higher polarization rate and polarization angle. However, both { polarization vectors are nearly  perpendicular to the observed radio jet projection.}  It indicates that the polarization in the continuum is coming from the accretion disc. Our simple lensing model  of a polarized source showed that in principle macro-lens can cause the observed differences in polarization parameters of Q0957+561 A,B images. Using Mg II broad line and luminosity of component A we estimated that the Q0957+561 black hole mass is $M_{\rm SMBH}\approx(4.8-6.1)\cdot 10^8M\odot$.}
   {}

   \keywords{gravitational lensing: strong  -- (galaxies:) quasars: individual: Q0957+561 -- emission lines -- polarization - supermassive black holes}

   \maketitle
%

\section{Introduction}

Gravitationally lensed quasars are very important for a number of investigations in astrophysics. First of all, the light from these objects is amplified and we can detect  objects at large redshift, therefore investigation  of lensed quasar and their geometry (in combination with the foreground galaxy) is important for cosmology. Additionally, gravitational lenses can be used to constrain the innermost structure of lensed quasars \citep[see e.g.][]{ji14,br17,hu17,pa20}, which are  a specific class of the active galactic nuclei (AGNs).  { Different emitting regions (which are of different dimensions) of a lensed quasar can be differently affected by microlensing  \citep[see e.g.][]{jov08}. This can cause chromatic effects \citep[][]{pop05} in an image of the lensed quasar spectrum. Therefore,    variations in the spectral  characteristics  of an image  of lensed quasars can constrain the quasar inner structure   \citep[see e.g.][etc.]{pop01,ab02,pop05,ab07,sl07,bl11,fi16,fi18}. For example,}   gravitational microlensing {   effect} gives a possibility to explore the  accretion disc structure and its temperature profile \citep[see e.g.][]{co20}, but also the structure and kinematics of the
 broad line region  \citep[BLR, see e.g.][]{pop01,ab02,sl12,gu13,br17,hu17}, which emits broad lines. 

 The broad lines are originated relatively   close to the central  supermassive black hole (SMBH), assumed to be in the center of AGNs,  and  can be used to measure its mass
\citep[see][]{pet14,med18,med19,pop20}. There is a possibility that the  BLR emission is amplified by the microlensing effect, and consequently, the  broad lines can be affected by this effect
 \citep[see e.g.][etc]{pop01,ab02,br17,hu17,fi18,pa20}. This may give information about the BLR dimension and kinematics. 
 
 The BLR structure and kinematics  can be investigated  by using polarization characteristics across a broad line profile \citep[see e.g.][etc.]{sm04,af14,sa18,af18}. As it was shown recently by \citet[][]{af18}, polarization in  broad lines can be used to explore the BLR kinematics, inclination and dimensions. Moreover, another important finding is that the polarization in broad lines can be used for SMBH mass estimates \citep[see][]{ap15,sa18,af18}. 

Consequently, spectropolarimetric observations can give very useful information about the structure of lensed quasars \citep[see e.g.][etc.]{hu98,bl00,hl07,hu15,pa20}. However, the nature of polarization in lensed quasars is not yet clearly understood. Comparing different images of SDSS J1004+4112 lensed quasar, \citet[][]{pa20} found that significant change in polarization parameters (observed only in the component D) can be explained by microlensing of  a scattering region located in the inner part of a dusty torus. Different mechanisms could contribute to the polarization in quasars \citep[][]{sm04}, and different effects in the lensed polarized light can be expected. In order to continue our investigation of spectropolarimetric characteristics of lensed quasars, we observed the lensed quasar Q0957+561 with the 6m telescope of Special Astrophysical Observatory (SAO RAS) in the polarization and spectral modes.

The first identified gravitationally lensed quasar Q0957+561 \citep{wa79} has two images of a quasar with  redshift $z = 1.41$ that is lensed by a foreground bright galaxy at  $z  =  0.36$ in a cluster of galaxies \citep[see e.g.][]{go80,young80,young81,rh91,ke00}. There are two images A and B of a lensed quasar projected at the distance of more than 6$^{\prime\prime}$. The spectra of both images show broad emission lines \citep[see e.g.][]{wa79,young81}, which are usually observed in the spectrum of a Type 1 quasars. The images A and B have been observed in X-ray \citep[see e.g.][]{ch95} and radio \citep[see e.g.][]{gr85,ga94,ca95,re95,ha97,ha08}  spectral bands.
A time delay between components A and B is around 420-425 days  \citep[see e.g.][etc.]{sch90,bo95,pi97,ku97,oscoz01,ovaldsen03,sha08,sha12}. The images of Q0957+561 quasar have been monitored in different spectral bands \citep[see, e.g.][]{ch95,ca95,go08}, and  investigations of the innermost structure are performed \citep[see e.g.][]{sch05,ha12} using  variability of the images. Also, polarization in both images has been reported by \cite{do95}.

  {In \cite{pa20} we explored the polarization of  a radio-quite lensed quasar SDSS J1004+4112 and found  significant differences between polarization parameters in different images. But, due to somewhat lower  brightness of lens components, we could not explore the polarization across the broad line profiles. To explore the polarization in broad lines of a gravitational lens, one has to select lens with enough bright and separated images.} This motivates us to  observe the lensed  radio-loud quasar Q0957+561 in the spectroscopic and polarization mode with the 6m telescope of SAO observatory.

 We obtained spectroscopic and polarimetric observations of Q0957+561 in February and April  { 2020}. The idea was to explore spectral and polarization characteristics of the lensed quasar and find the possible influence of macro-lensing to the polarization of lensed quasars.
 
{ Throughout this paper, we adopt the  following cosmological parameters: $\Omega_m = 0.27, \Omega_\Lambda = 0.73$ and $H_0 = 71$ km/s/Mpc }
 The paper is organized as  follows:
 In   \S2 we describe our observations, in \S 3 we give the results, which are discussed  in    \S 4.   The main conclusions are summarized  in \S5.
 
 \

\section{Observations and  Data Reduction}\label{obs}

The first gravitational quasar lens Q0957+561 was observed in the spring of 2020 with the 6-m telescope using the universal spectrograph SCORPIO-2 in various modes \citep{am11}. Initially, the task was to study the polarization in broad lines using spectropolarimetric data and  measure the mass of the central black hole \citep[see][]{ap15}. But, obtained polarization angle shape across broad Mg II line profile { indicates that the equatorial scattering mechanism   \citep[typical for Type 1 AGNs, see][]{sm04,af18} is not dominant.} Therefore we performed additional spectral observations in non-polarized light with a high signal-to-noise ratio and high-precision photometry and polarimetry of Q0957+561A,B images.

\subsection{Spectropolarimetry}\label{spol}

For spectropolarimetric observations with SCORPIO-2, we used a double Wollaston prism as an analyzer in a parallel beam of a focal reducer. In such analyzer, the beam divided into two halves enters two Wollaston prisms separating the directions of the polarization plane   $0^\circ$-$90^\circ$ and $45^\circ$-$135^\circ$, respectively. It allows  to register simultaneously four spectra of an object in four polarization planes and determine the Stokes parameters based on these data. On February 16, 2020, in this mode the observations of Q0957+561 were made under good atmospheric conditions (seeing 1.2\arcsec~ and variations of the polarization channels transmission of $<$1\%). The slit width was 2\arcsec~ and its length was 60\arcsec. The slit passed through both images of the lensed quasar (position angle $168^\circ$). A series of quasar spectra were obtained with a total exposure of 3900 sec with a spectral resolution of 14\AA~ in the range of 4200-7400\AA\AA~ (VPHG940{\tt @}600 grating). Images were registered on the EEV42-90 CCD with the format $4096 \times 2048$ px \citep{mu16}. On the same night, spectra of the spectropolarimetric standards (G191B2B and BD+59d389 stars) were obtained at close zenith distances to calibrate the transmission of the polarization channels. The technique of polarization observations, calibration, and data reduction are described in  \citet{am12}. As a result of the reduction, the spectra of components A and B of the gravitational lens were obtained in four directions of polarization $I_0(\lambda), I_{90}(\lambda), I_{45}(\lambda)$ and $I_{135}(\lambda)$. In this case, the first three Stokes parameters can be found from the relations:
\begin{equation}
\begin{array}{c}
I(\lambda)=I_0(\lambda)+I_{90}(\lambda)K_Q(\lambda)+I_{45}(\lambda)+I_{135}(\lambda)K_U(\lambda)\\
\end{array}
\end{equation}
\begin{equation}
\begin{array}{c}
Q(\lambda)=\displaystyle{\frac{I_{0}(\lambda)-I_{90}(\lambda)K_Q(\lambda)}
{I_{0}(\lambda)+I_{90}(\lambda)K_Q(\lambda)}}\\
\end{array}
\end{equation}
\begin{equation}
\begin{array}{c}
U(\lambda)=\displaystyle{\frac{I_{45}(\lambda)-I_{135}(\lambda)K_U(\lambda)}
{I_{45}(\lambda)+I_{135}(\lambda)K_U(\lambda)}}\\
\end{array}
\end{equation}
Here $K_Q(\lambda)$ and $K_U(\lambda)$ are instrumental parameters that define the dependence of the transmission of spectral channels on the wavelength specified by observations of zero polarization standard stars.

\subsection{Spectrophotometry }

Spectra with a high signal-to-noise ratio were obtained on April 19, 2020, by the SCORPIO-2 spectrograph in a long-slit mode  in the range 3700-7300\AA~ using VPHG1200{\tt @}540  grating with good atmospheric transparency and  ~1.2\arcsec  seeing. The six spectra were obtained with a total exposure of 1800 sec. The spectral resolution determined from the lines of the night sky was 7.5\AA. We obtained the spectrum of the spectrophotometric standard BD+75d325 at a close zenith distance for absolute flux calibration. The new sensor CCD261-84 $2048 \times 4096$ px with a size of 15 microns was used. This CCD is manufactured by new 'high-rho' technology used to increase the thickness of the silicon to maximize the response at the infra-red end of the spectral range. Such a device, except for its high quantum efficiency ($>$90\% at 400-900 nm and $>$40\% at 350 and 1000 nm) in the entire visible range, has practically no fringes (their amplitude is $<$0.2\%) in the red region \citep[see][]{jor10}. A special feature of the device is a large number of cosmic ray hits registered even at short exposures, which creates difficulties in data reduction.  

Data reduction was carried out using the standard method  for reducing long-slit spectra: construction of a two-dimensional geometric distortion model followed by bi-linear interpolation of 2D spectra into a rectangular  coordinate grid, linearization and correction of the spectral flat-field, subtraction of the sky background, removing cosmic ray hints and extraction of spectra \citep{am12}.

\subsection{Photometry and polarimetry }

\begin{figure}[]
\centering
\includegraphics[width=9 cm]{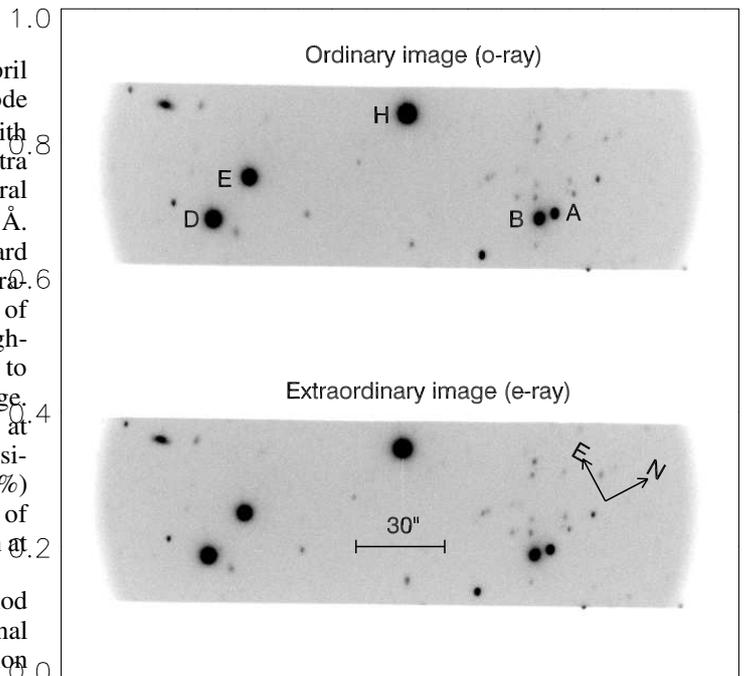}
\caption{Comparison of ordinary (o-ray) and extraordinary (e-ray) images.}
\label{fig01a}
\end{figure}

Image polarimetry of Q0957+561A,B components was performed on April 24, 2020, with the SCORPIO-2 spectrograph in the g-SDSS and r-SDSS filters. The Wollaston prism was used as an analyzer in combination with a rotating half-wave phase plate. The plate rotates at four fixed angles: 0$^\circ$, 22.5$^\circ$, 45$^\circ$  and 67.5$^\circ$.  For each angle, we registered two 3\arcmin\ $\times$ 2\arcmin\ images on CCD261-84 in two polarization directions 0$^\circ$ (o-ray) and 90$^\circ$ (e-ray), as shown in Fig. \ref{fig01a}. 32  images (8 series of exposures for the four positions of the phase plate) were obtained with an exposure of 200 seconds. For ordinary $F_o$ and extraordinary $F_e$ rays, the fluxes of the studied objects were measured using aperture photometry on each frame. Then the dimensionless values $F=(F_o-F_e)/(F_o+F_e)$ were calculated. The dimensionless Stokes parameters $Q$ and $U$ can be found from the known relation:
\begin{equation}
\begin{array}{c}
	Q = \frac{1}{2}  \left( \left(\frac{F_o-F_e}{F_o+F_e } \right) _{\theta=0^\circ } - \left( \frac{F_o-F_e}{F_o+F_e }\right) _{\theta=45^\circ}  \right),
\end{array}
\end{equation}
\begin{equation}
\begin{array}{c}
	U = \frac{1}{2}  \left( \left(\frac{F_o-F_e}{F_o+F_e } \right) _{\theta=22.5^\circ } - \left( \frac{F_o-F_e}{F_o+F_e }\right) _{\theta=67.5^\circ}  \right),  
\end{array}
\end{equation}
where $\theta$ is the rotation angle of the phase plate.

The variations of the atmospheric transparency during the observations did not exceed 0.5\%, and seeing was 1.4-1.7\arcsec.  Images of the zero polarization standard BD+32d3739 were also obtained to determine instrumental polarization and the highly polarized standard Hiltner 960 was observed to control the direction of the polarization angle. The secondary standards in the Q0957+561 field (stars D, E and H, as it is shown in Fig.\ref{fig01a}) taken from \citet{ovaldsen03} were used for the photometric binding. The magnitudes of these stars in the g-SDSS and r–SDSS filters are taken from \citet{sha08}. In individual frames, the signal-to-noise ratio is 300-400 for quasar and $>$2000 for the referent stars.  However, transparency variations can worsen the photometric accuracy, and for further processing, we used the method of differential photometry and polarimetry, described in details in  \citep{ShAf19}.  To determine the flux, we added the measurements obtained from the two directions of polarization. Thus, 32 independent measurements were made for each object in each filter.  Table \ref{photometry} shows the results of photometry of the Q0957+561A,B components and two reference stars D and E relative to the H star.

\begin{table*}[]
\begin{center}
\caption{Photometry of Q0957+561.}
\label{photometry}
\begin{tabular}{|l|l|l|l|}
\hline
Object & g-SDSS                  & r-SDSS                 & (g-r)       \\ \hline
Star D & 15.481$\pm$0.002 (15.486)   & 14.946$\pm$0.003  (14.951) & 0.535$\pm$0.004 \\
Star E & 15.858$\pm$0.003   (15.816) & 15.240$\pm$0.002  (15.217) & 0.599$\pm$0.005 \\ \hline
QSO A  & 17.825$\pm$0.004            & 17.504$\pm$0.004           & 0.321$\pm$0.006 \\
QSO B  & 17.159$\pm$0.002            & 16.895$\pm$0.003           & 0.264$\pm$0.004 \\ \hline
\end{tabular}
\end{center}
\end{table*}

\section{Results}

\subsection{Spectral characteristics of Q0957+561 A and B images}

\begin{figure}[]
\centering
\includegraphics[width=9 cm]{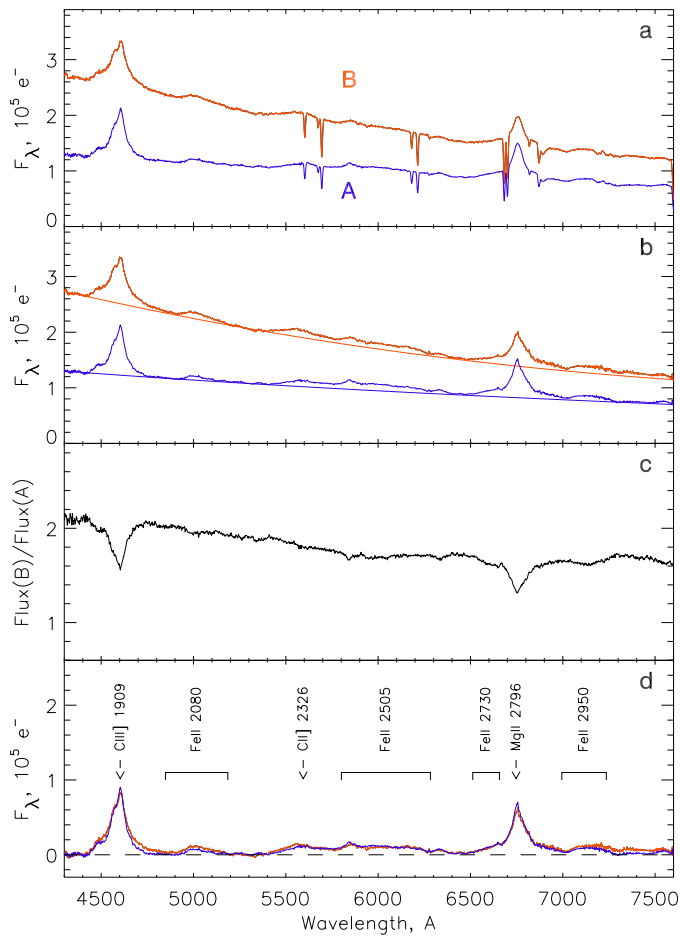}
\caption{Integral spectra of components A and B of the lensed quasar Q0957+561: a) the observed A and B spectra; b) the A and B spectra { without the absorption lines}; c) the B/A flux ratio as a function of wavelength; d) the emission line spectra of the A and B components after subtracting the continuum. The lines are identified according to \citet{bo90}.}
\label{fig01}
\end{figure}

\begin{figure}[h]
\centering
\includegraphics[width=9cm]{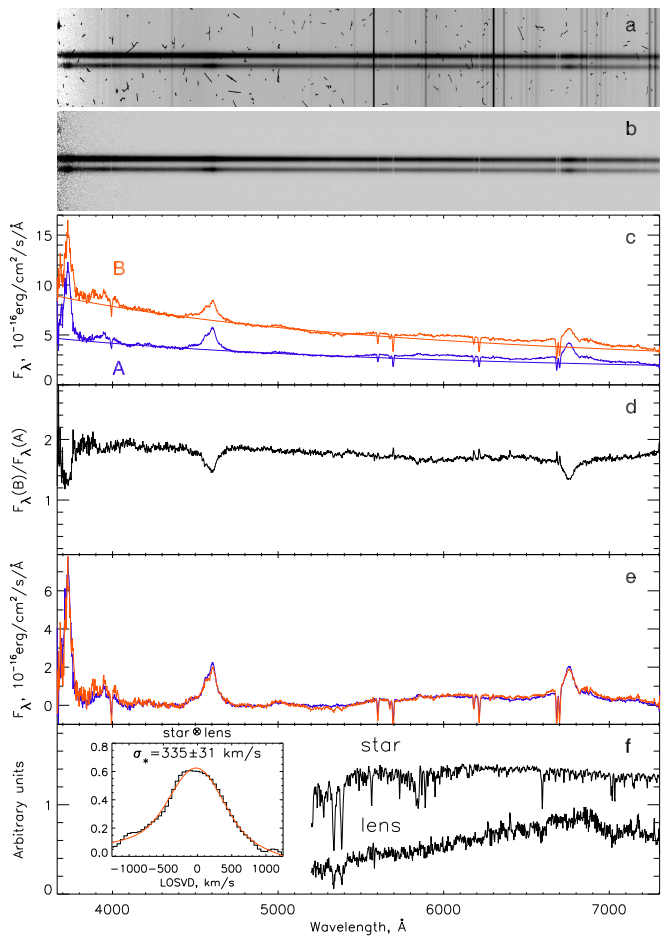}
\caption{The long-slit spectra of Q0957+561A,B components: a) the linearized source spectrum, b) the sky subtracted spectrum with removed cosmic rays; c) the extracted spectra of components A (blue) and B (red) of the lensed quasar image; d) the brightness ratio of the components; e) the spectra of components A (blue) and B (red) { after continuum subtraction}; f) the spectrum of the lensing galaxy in arbitrary units and the comparison star, shifted by $z = 0.36$. The bottom panel (left) shows the cross-correlation function between the spectra of the star and lens galaxy.}
\label{fig02}
\end{figure}

As can be seen from  Table \ref{photometry}, the brightness ratio of the B/A components is 1.85 in the blue region of the spectrum, and 1.75 in the red one. This can be clearly seen in Figs. \ref{fig01} and \ref{fig02}.

Figure \ref{fig01}a shows the integral spectra $I(\lambda)$ calculated according to Eq. (1) for components A and B and corrected for spectral sensitivity. The spectrum of each component contains the strong broad emission lines CIII] 1909\AA, Mg II 2790\AA, and weak blended lines of Fe II multiplets. In the Mg II region, there are narrow metal lines  belonging to the quasar and an atmospheric absorption band O$_2$ which make  difficult the analysis of the polarization in broad lines and require that  these lines are first removed from the spectrum. Fig. \ref{fig01}b shows the spectra with absorption lines removed.

Our spectra (Fig. \ref{fig01}c) clearly show that the component B (south) is more than 1.5 times brighter than the component A (north), although the brightness ratio of B/A was less than 1 when the  Q0957+561 lens system was discovered by \cite{wa79}. This ratio changes with the wavelength.  \cite{wa79} showed the increase of the B/A ratio with the wavelength,  but our observations show that the ratio decreases with the wavelength. As { can be seen in} Fig. \ref{fig01}c, the brightness ratio of components B and A in the emission lines is less by 15-20\% than in the continuum. To understand this, we subtract a continuum approximated by a power-law dependency $\lambda^{-\alpha}$. 
Estimations of $\alpha$ for the components B and A are of $1.43\pm0.10$ and $1.07\pm0.09$, respectively. 
The emission spectrum of both components after subtracting the continuum is presented in Fig. \ref{fig01}d. As can be seen, the component spectra match up to errors, which means that in the emission lines, the gravitational brightness amplification is not observed in the lens components of Q0957+561. To verify this observing fact, we performed additional spectral observations with a high signal-to-noise ratio, and the results  of these observations are shown in Fig. \ref{fig02}.

Fig. \ref{fig02}a shows the image of the original linearized spectrum and Fig. \ref{fig02}b shows the spectrum after subtracting the background and removing cosmic rays. The correction for atmospheric extinction for the spectrophotometric standard and the object was performed in a standard way, taking into account measurements of the spectral transparency of the atmosphere at the 6m telescope location \citep{ka78}.

The spectra of Q0957+561 components corrected for spectral sensitivity are presented in Fig. \ref{fig02}c. Atmospheric absorption in O$_2$ band is accounted by observations of a bright star in the field.  {The difference in the slope of the component spectra, detected in our  first observation (spectropolarimetric mode), can be  seen quite confidently. The indexes $\alpha$ for components B and A are equal to 1.41$\pm$0.07 and 1.16$\pm$0.15, which corresponds to the spectropolarimetric measurements within the error limits.} The ratio $F_{\lambda}(B)/F_{\lambda}(A)$ shown in Fig. \ref{fig02}d also corresponds to the one obtained by spectropolarimetry. The difference of about 5\% is due to different image qualities when during observations. Fig. \ref{fig02}e shows the spectra of both components after subtracting the power-law continuum. The spectra do not reveal any significant difference in the fluxes of emission lines, which confirms the result obtained by spectropolarimetry.  In the spectrum of component B, a small increase in intensity is seen in the region of 6000-7000\AA~ due to a lensing galaxy entering the slit located in 0.6\arcsec~ in the projection. Fig. \ref{fig02}f shows the spectrum of the lensing galaxy as a result of subtracting the spectra of the components taking into account the difference of their brightness. The same figure shows the spectrum of the HD245 star of the G2V spectral class, taken from the MILES spectrum library \citep{sa06}. 
The star spectrum is shifted to $z=0.36$.  The cross-correlation analysis of the star and lensing galaxy spectra shown in the rest frame and corrected for spectral resolution is plotted on the left in Fig. \ref{fig02}f. The Gaussian approximation of the cross-correlation function gives an estimate of the dispersion of stellar velocities in the lens of $\sigma_*=335\pm$31 km s$^{-1}$. This is in an agreement  with the measurements reported by \citet{med00}. They obtained the central stellar dispersion  of   $\sigma_*=310\pm$20 km s$^{-1}$ for the lens galaxy G1 associated with Q0957+561A,B. 

\begin{figure}[h]
\centering
\includegraphics[width=9 cm]{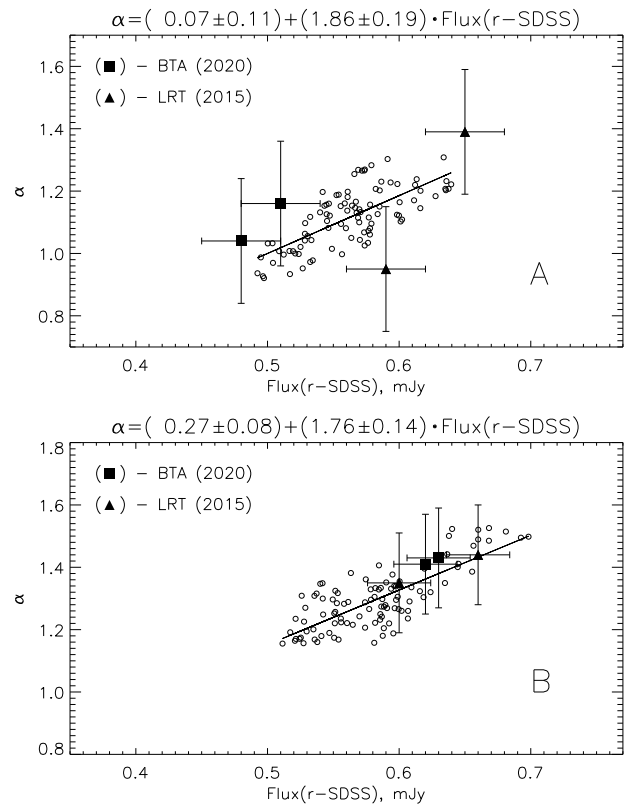}
\caption{Changes of the slope $\alpha$  as a function of the observed flux for component A (upper panel) and B (bottom panel).  {Open circles denote the photometric data taken from \citet{sha12}.} Full squares denote our observations (BTA), and {   full} triangles the observations with LRT given in the database of bright quasars \citep[see][]{gi18}.  {The BTA and LRT observations were conducted in the spectral mode which explains the significant error-bars.}}
\label{fig03}
\end{figure}

We explore  B/A ratio from previous observations and found that, when the lens was discovered, the initial brightness ratio B/A was 0.76 in the blue part and 1 in the red part \citep{wa79}. In the 1980s, \citet{go80} observed the B/A ratio  around 0.7 in the whole spectral range, and according to \cite{va89}, the B/A ratio in 1980-1983 was between 0.85 and 0.95 in the red part, while in the blue part the B/A ratio stays constant on the level of  0.5. Measurements of B/A ratio in papers from 1990s to 2008 show that the B component stays brighter (the ratio B/A$>$1), i.e.  the values B/A were between 1.05 and 1.22 \citep[see][]{sch90,co02,co03,sha08}. It is obvious that the B/A ratio is changing, and consequently, one can expect that the power-law index $\alpha$ is changing during time. To explore the changes of index $\alpha$ for components A and B, we use a {spectral database\footnote{https://grupos.unican.es/glendama/database/} of bright lensed quasars \citep[][]{gi18}}.  We use observations from the Liverpool Robotic Telescope (LRT) obtained in 2015 and our observations in 2020.  {Furthermore, to investigate the behaviour of $\alpha$ we added the photometric data from \citet{sha12} corrected for a $\sim$417 days delay between the lens components. The relation between the slope $\alpha$  and the flux of each component is plotted in Fig. \ref{fig03}. } 

As can be seen in Fig. \ref{fig03}, the slope $\alpha$  is well  correlated with the flux for both components. It indicates that in the brighter phase  the blue part of spectra is amplified.

\subsection{Polarization   of Q0957+561 A and B images}

We found the Stokes parameters $Q(\lambda)$ and $U(\lambda)$, and then calculated  the polarization degree $P(\lambda)$ and the angle of the polarization plane $\varphi(\lambda)$ as functions of wavelength using the known relations:
\begin{equation}
\begin{array}{ccc}
P(\lambda)=\sqrt{Q(\lambda)^2+U(\lambda)^2,} \\ \\ \varphi(\lambda)={1\over2}\arctan{[{U(\lambda)}/{Q(\lambda)]}}+\varphi_0~,
\end{array}
\end{equation}
where $\varphi_0$ is the zero point determined by observations of highly polarized standard star.  {The $\pi/2$ ambiguity of the polarization angle is corrected according to the formulae given in \cite{Bag09}.}

\begin{figure*}[t]
\centering
\includegraphics[width=16 cm]{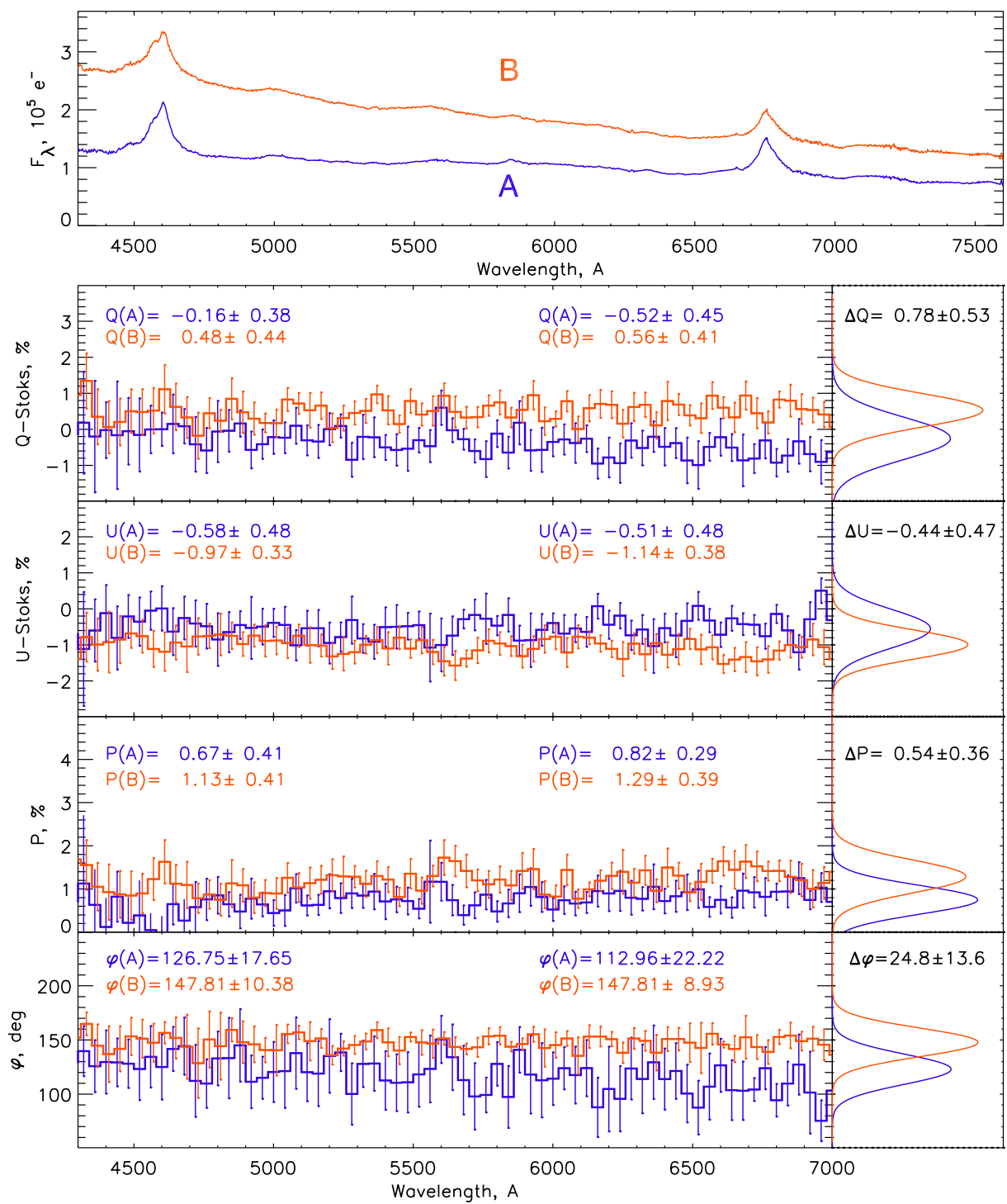}
\caption{Spectra of A and B components (first panel) and their polarization parameters as function of wavelength (2-5 panels). Observed Stokes parameters (2nd and 3rd panels), the degree of polarization (4th panel), and the angle of the polarization plane (5th panel) for components A (blue) and B (red) of Q0957+561 lens are shown. {The spectropolarimetric data on 2-5 panels are binned over 40\AA.} The left sides of 2-5 panels show robust estimates for each parameter in the g-SDSS (left) and r-SDSS (right) bands. The right sides of 2-5 panels show the distribution of parameter values and their average difference for B-A.}
\label{fig04}
\end{figure*}

As it can be seen in Fig. \ref{fig04} the polarization parameters seem to be different for different components, and we could not find the expected 'S' shaped profile in broad lines, especially in the Mg II broad line.

 The first panel in Fig. \ref{fig04} shows the integral spectra of A and B components given for comparison with the polarized spectra  presented on panels 2-5. On each of 2-5 panels,  the average robust estimates of the polarization parameters
 and their errors for the g-SDSS (left) and r-SDSS bands (right) are given. 
 The given errors are the robust standard deviation. 
From top bottom  we plot Stokes parameters Q (2nd panel), U (3rd panel), and polarization parameters: polarization degree in percents (4th panel) and polarization angle (5th panel). The data for the B component are denoted in red and the fro the  A component  in blue. All quantities are given as functions of wavelengths.

On the right sides of 2-5 panels  (Fig. \ref{fig04}) the averaged values and distribution of polarization parameters are given. The true accuracy of the estimates of the Stokes parameters and the degree of polarization is about $0.5\div0.7\%$, and the accuracy of the angle estimation is $15^\circ\div25^\circ$. The accuracy is affected not only by the value of the measured flux in the polarization channels but also by variations of the atmospheric depolarization at different exposures and errors of integration of the component spectra. As it can be seen in Fig. \ref{fig04}, the accuracy of measuring the polarization parameters is slightly higher for the brighter component B than for A. 
 { However, as one can see on the right panels in Fig. \ref{fig04} the statistical difference between the polarization parameter distributions  of A and B components is present.}

Table \ref{polarimetry} shows the results of measuring the Q0957+561 polarization parameters using image-polarimetry data. There are  estimates based on broad-band spectropolarimetry data obtained by integration in the g-SDSS and r-SDSS bands. The accuracy of the image polarimetry is $\sim$0.1-0.2\% for the polarization degree and ~2-3$^\circ$ for the polarization angle.

The polarimetry data confirm the $\sim$1.5 times difference in the degree of polarization of components A and B detected by spectropolarimetry.  The degree of polarization for each component does not depend on the wavelength within the error-bars. The polarization angle of component B is the same within the error-bars  in the blue and red parts of the spectrum and is equal to $\sim$150$^\circ$. For component A, the polarization angle changes along the spectrum: from 130$^\circ$ in the blue region to $\sim$120$^\circ$ in the red one.

\begin{table*}[]
\begin{center}
\caption{Polarimetry of Q0957+561.}
\label{polarimetry}
\begin{tabular}{|l|l|l|l|l|l|}
\hline
\multicolumn{1}{|c|}{Filter} & \multicolumn{1}{c|}{} & \multicolumn{1}{c|}{$Q$-Stokes,\%} & \multicolumn{1}{c|}{$U$-Stokes,   \%} & \multicolumn{1}{c|}{P, \%} & \multicolumn{1}{c|}{$\varphi$, deg} \\ \hline
\multicolumn{6}{|c|}{Spectropolarimetry, 16 Feb 2020}                                                                                                                                                \\ \hline
\multirow{2}{*}{g-SDSS}      & A  & -0.16$\pm$0.38    & -0.58$\pm$0.48   & 0.67$\pm$0.41   & 127$\pm$18                            \\ \cline{2-6} 
                             & B  & 0.48$\pm$0.44     & -0.97$\pm$0.33   & 1.13$\pm$0.41   & 148$\pm$10                            \\ \hline
\multirow{2}{*}{r-SDSS}      & A  & -0.52$\pm$0.45    & -0.51$\pm$0.48  & 0.82$\pm$0.29   & 113$\pm$22                            \\ \cline{2-6} 
                             & B  & 0.56$\pm$0.41   & -1.14$\pm$0.38    & 1.29$\pm$0.39   & 148$\pm$9                             \\ \hline
\multicolumn{6}{|c|}{Image polarimetry, 24 Apr 2020}                                                                                                                                                 \\ \hline
\multirow{2}{*}{g-SDSS}      & A & -0.16$\pm$0.20  & -0.78$\pm$0.14    & 0.75$\pm$0.08   & 130$\pm$3                             \\ \cline{2-6} 
                             & B & 0.65$\pm$0.17   & -0.81$\pm$0.25   & 1.04$\pm$0.07    & 153$\pm$2                             \\ \hline
\multirow{2}{*}{r-SDSS}      & A & -0.39$\pm$0.06   & -0.52$\pm$0.09 & 0.67$\pm$0.09     & 117$\pm$3                             \\ \cline{2-6} 
                             & B & 0.61$\pm$0.06    & -0.99$\pm$0.09 & 1.16$\pm$0.05     & 151$\pm$2                             \\ \hline
\end{tabular}
\end{center}
\end{table*}

The spectra of the both components (see e.g. Fig.\ref{fig01}) show the broad lines, e.g. Mg II, which may come from the partly  virialized BLR and in principle can be used for mass determination \citep[for review see][and reference therein]{pop20}. In \cite{sa20}  we show that in the case of Mg II line, where outflows/inflows in far wings can be present, the 'S' shaped polarization angle profile  across the broad line can still be present.  Yet, the features 'S' shaped polarization angle across Mg II   are not present in components A and B.
As can be seen in Fig. \ref{fig04}, comparing the integral spectra with the polarization,  we can conclude that Q0957+561A,B components have no significant changes in the polarization angle across the  broad emission lines within the measured error-bars. It {   may} indicate that the polarization mechanism is probably not related to equatorial scattering on the dusty torus \citep[see][]{sa18,sa20}.

\begin{figure}[]
\includegraphics[width=9cm]{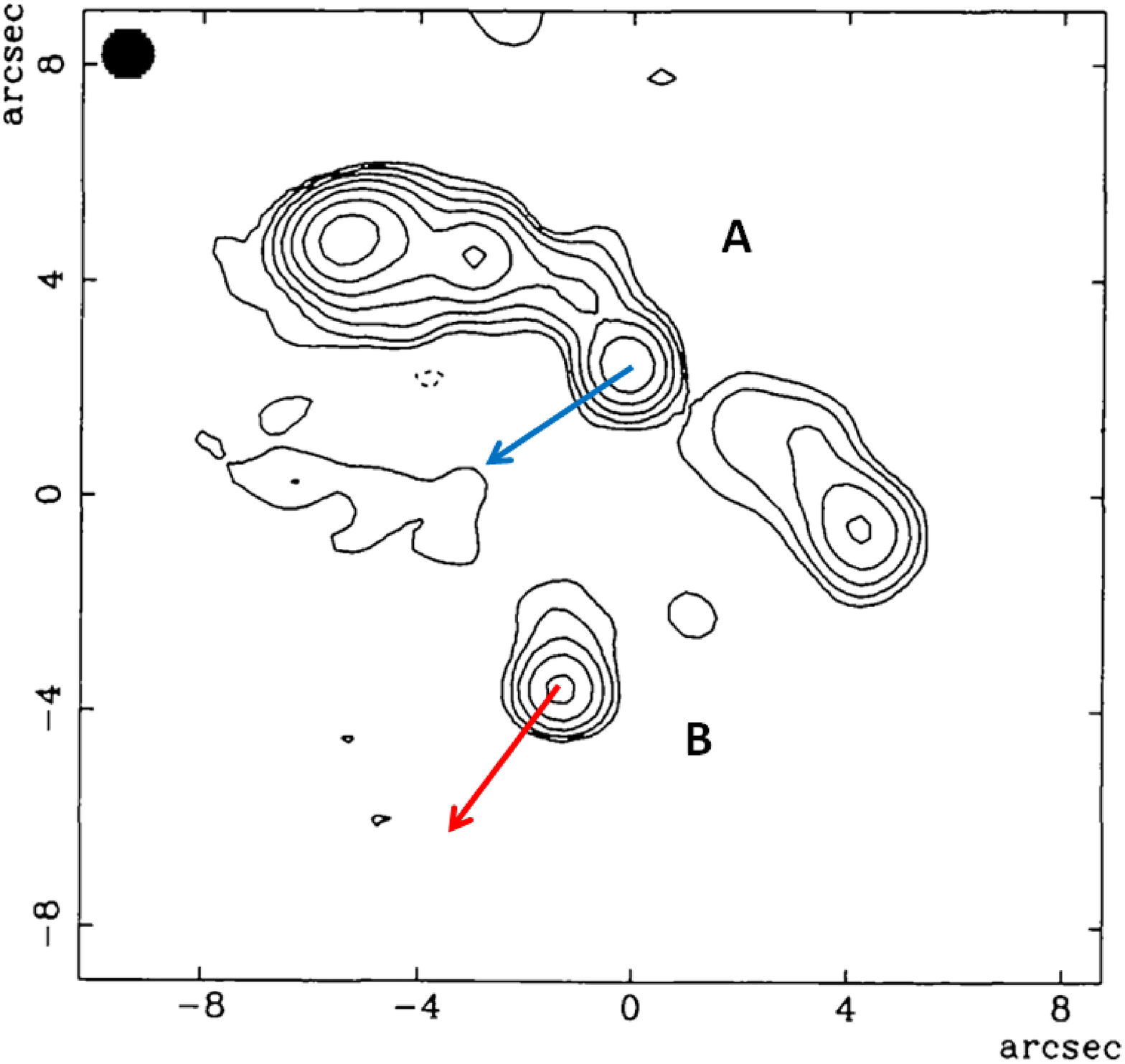}
\caption{The orientation of the polarization vectors in the A and B component of  Q0957+561  gravitational lens. The vectors { (as arrows)} are over-plotted on the composite radio image of the lens   taken from \citet{re95}.}
\label{fig-pol}
\end{figure}

The polarization of the continuum has a direction of the electric vector between $120^\circ$ (component A) and $150^\circ$ (component B) which is approximately perpendicular to the radio jet axis (see Fig. \ref{fig-pol}). { In Fig.   \ref{fig-pol} we over-plot the polarization vector  (shown as  arrows) on the composite radio-image of the Q0957+561 lens system taken from \citet{re95}.  The polarization vector  is almost perpendicular to the radio-jet observed in component A. Taking that   $\lambda$18cm global VLBI hybrid maps  \citep[see Fig. 2 in][]{ga94}  show nearly parallel radio-jets of A and B components on the {\it mas} scale, it seems that the polarization vector is nearly perpendicular to the radio-jet in the source.}

\subsection{SMBH mass  of Q0957+561}

{    We were not able to measure Q0957+561 SMBH mass using the  polarization in the broad   lines  caused by equatorial scattering.} However, a combination of spectral and photometric observations allowed us to estimate the absolute values of the emitted flux and, consequently, the mass of the central SMBH. 

To do this, we measured the flux of component A, which is further away from the lensing galaxy and probably is not microlensed, at a wavelength 3000\AA\ in the reference frame of the lensed quasar: $F_{\lambda} = (3.2 \pm 0.6) \times 10^{-16}$ erg cm$^{-2}$ s$^{-1}$ \AA$^{-1}$. 
The amplified quasar luminosity is obtained as $\mu\times(\lambda L_{3000}) =   (1.2 \pm 0.3) \times 10^{46}$ erg cm$^{-2}$, where $\mu$ is the amplification of component A due to macrolensing.

The amplification estimation was done similarly as in \citet{pa20}, taking $\kappa \approx 0.47$  \citep{nakaj} and $\gamma \approx 0.1$  \citep{fadely,krisp} for component A, we obtained $\mu = 3.7$. The size of the BLR region in the Mg II line is estimated using the   empirical  BLR radius - luminosity (R-L) relation \citep[see][]{bozena,pop20}. We used an updated R-L  (at 3000 \AA) relation given by  \citet{zaja} and obtained $R_{\rm BLR, Mg II} = 227^{+104}_{-72} $ light days. 

After subtracting the Fe II contribution to the Mg II line using  the UV Fe II model given in  \citet{pop19}\footnote{models of the UV Fe II can be found at \url{http://servo.aob.rs/FeII_AGN/link7.html}} we  measured the FWHM =$3.68 \times 10^3 \ {\rm km\  s^{-1}}$ and estimated cloud velocity as $\sigma ={\rm FWHM/2.355}= 1.58  \times 10^3 \ {\rm km\  s^{-1}}$.

 Then using the relation \citep[see][]{pet14}:
\begin{equation}
\label{formula:virial1}
M_{\rm SMBH} = f\cdot (R \sigma^2 G^{-1}),
\end{equation}
where $f$ is a dimensionless parameter equal to 5.5 \citep{onken} depending on the BLR structure and kinematics and the inclination of the system relative to the observer, $ G $ is the gravitational constant, $R$  is the BLR dimension. We estimated that SMBH mass is $M_{\rm SMBH} \approx 6.1 \times 10^8 M_{\odot}$.

Also we estimated SMBH mass directly using $\lambda L_{3000}$ and FWHM of the broad Mg II emission line, according to the relation given in \citet{pop20}:
\begin{equation}
\label{formula:virial2}
\log(M_{\rm SMBH}) = 1.15 + 0.46\cdot\log(\lambda L_{3000}) \\ + 1.48\cdot\log({\rm  FWHM}),
\end{equation}
where $\lambda L_{3000}$ is given in units of $10^{44}$ erg s$^{-1}$ and  FWHM is in units of $10^{3} {\rm \ km \ s^{-1}}$. Using this relationship we found that the SMBH mass is $M_{\rm SMBH} = 4.8  \times 10^8 M_{\odot}$, which is close to the result obtained above. Our SMBH estimates $M_{\rm SMBH}\approx(4.8-6.1)\times 10^8M\odot$ are in agreement with \citet{as11}, where they obtained $(4.7-9.5)\times 10^8 M\odot$

 {We should note that we assume that the A component is not micro-lensed, however we cannot exclude this possibility, and the obtained SMBH mass may be affected by this effect.  }

\section{Discussion }

\subsection{Spectral characteristics: Changing in the innermost structure - intrinsic variability vs. microlensing}

{  The B/A flux ratio shows  the brighter component B than A}  in our observations and { also} in observations after 1990s.  However, in the epoch  of lensed quasar discovery  \citep{wa79}, and several years after that the component A was brighter in the UV spectra than the component B \citep{go80,va89}.  We also found that there is  a relationship between the change in the B/A ratio and slope $\alpha$. When  the images stay brighter, the blue part of their spectra stays more intensive (see Fig. \ref{fig03}). Also, \citet[][]{sha12} found that flux ratio oscillated in g-SDSS and r-SDSS bands during their observations, and that the ratio showed a slight increase during periods of the violent variability.

{ 
The variability, which causes the changes of spectral energy distribution (SED) in the UV band, is probably due to  perturbations of the inner quasar structure. It is known that  the temperature of the accretion disc can change due to  variations in the accretion rate \citep[see][]{ko99} and this will have an influence on the UV SED changes.  
 }

However, one cannot exclude the influence of microlensing on the observed changes in the UV SED. As it is well known, the microlensing effect is in principle achromatic, but if the dimensions of a disc (or a disc corona) are wavelength dependent \citep[i.e. the temperature is changing across the disc, see][]{jov08}, then one can expect that microlensing effect is chromatic \citep[see][]{pop05}, which will be observed as different amplifications in different wavelength bands \citep[see][]{jov08}.

One can expect an intrinsic variability that is dominant in component A, since the optical depth for microlensing of component A should be significantly smaller than of component B (component B is projected very close to the lens galaxy). And that extra variability detected in component B is due to microlensing. The amplification of the component B continuum seen in present days (that is brighter around two times than component A) is probably caused (at least partly)  by microlensing, since there is no line intensity variability (see Fig. \ref{fig01}, fourth panel) which is expected in the case of intrinsic variability. The line shape and their flux ratio is the same. Since the  BLR is significantly larger than the continuum source, the caustic can amplify the continuum source, but not the BLR emission \citep[see][]{ab02}.

The conclusion from the spectral observations is that the change in the A component spectra is mostly caused by the intrinsic variability of the quasar, but the variability in the component B is more complex, where both intrinsic variability and microlensing of the continuum source can contribute to the observed variability. The amplification of the component B is also reported in \citet{gi18}, where the B/A $>$1 was observed after 2011/12 (see their Fig. 6). Moreover, \citet{be19} found that  B has been microlensed in  recent epochs \citep[after 2011, similar as][]{gi18}. 

\subsection{Polarization of Q0957+561 A and B components}

\subsubsection{Polarization mechanisms in the lensed source}

{   We expected to observe polarization in Q0957+561A,B  broad line profiles which is typical for Type 1 AGNs, showing dominant equatorial scattering  \citep[see, e.g.][]{sm04,ap15,af18}}. 
However, as it can be seen in Fig. \ref{fig04} (5th panel), the so called 'S' shaped profiles of the polarization angle (PA) in broad CIV and Mg II broad lines {   is not present  \citep[as it is expected in Type 1 AGNs, see][]{sa18,sa20}.} 

The absence of 'S' shaped PA profile may indicate that: a) there is no equatorial scattering of the BLR light; b) the Keplerian motion is not dominant in the BLR. However, different BLR geometries produce different shapes in the polarization angle and different polarization degree in the broad lines. This is  also the case  of the  BLR with an outflow component \citep[see][]{sa20}, that is expected to be in the BLR emitting Mg II and C IV lines \citep[see][]{pop20etal,pop20}. However, as it can be seen in Fig. \ref{fig02} the polarization and PA in the broad lines is on the level of the continuum (within the error-bars). 

{ This indicates that some other effect can be present  as e.g. depolarization due to  a hot region located above the BLR. Similar effect is found  in 3C390.3 \citep[see][]{af15}, which is a radio-loud AGN.} 
This may be the case of Q0957+561, since the lensed quasar is a radio-loud object and there are  two radio components corresponding to the A and B images \citep[see][]{gr85,ro85}. Therefore, we can expect some outflow of hot gas above the BLR and significant depolarization.  Also, in the UV spectral region some kinds of outflow can be present, as e.g. \cite{sch05} indicated the bi-conical structures located  above and below the plane of the accretion disk, that is  apparently inclined for 55$^\circ$ to the line of sight.

 {An alternative explanation of the broad line polarization absence can be that the equatorial scattering is still present in the inner part of the torus. But, if the BLR dimension is comparable with the inner radius of torus the polarization in the broad lines cannot be detected \citep[see][]{ki04}. 
}

\begin{figure*}[t]
\includegraphics[width=9.5cm]{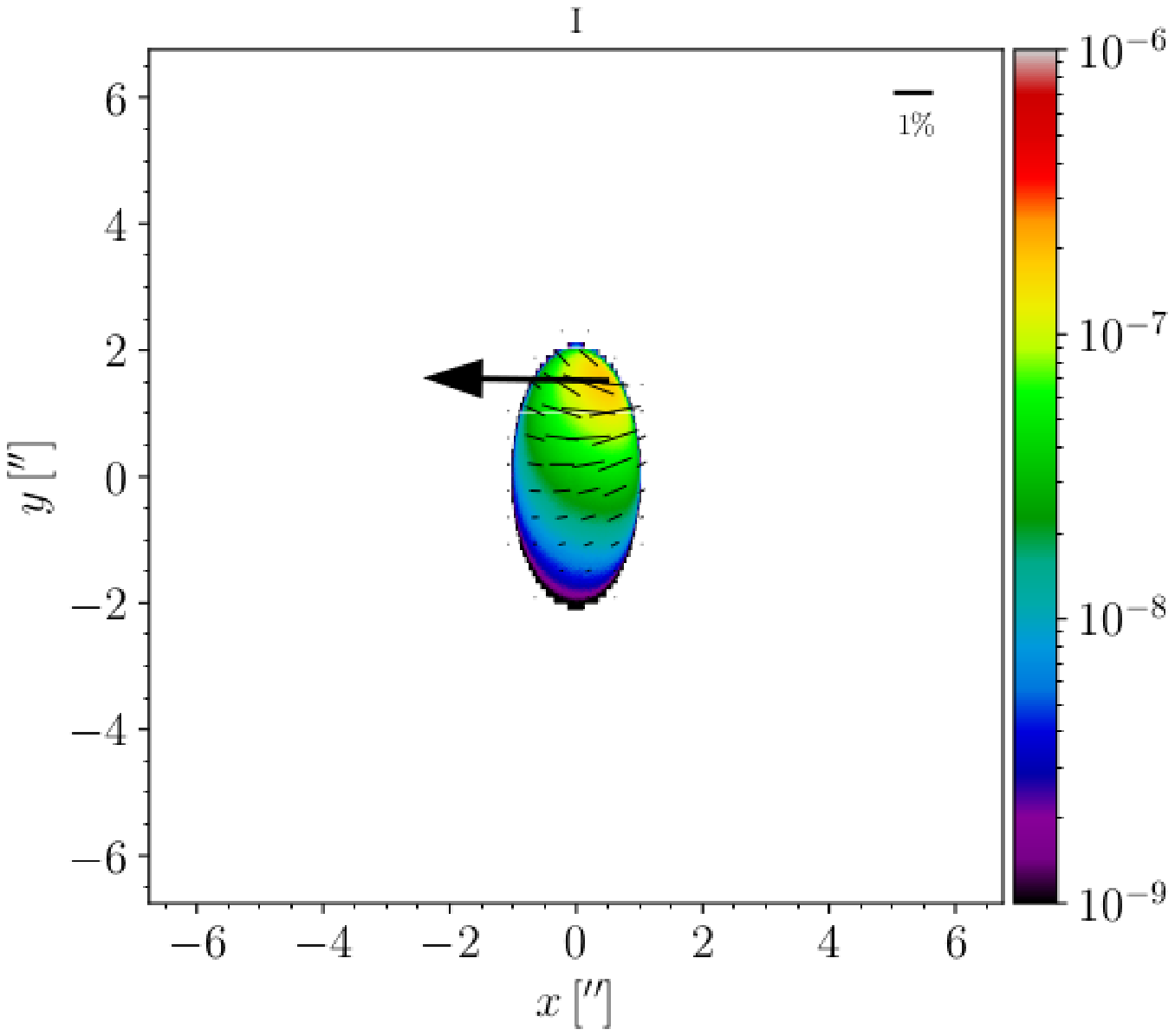}
\includegraphics[width=9.5cm]{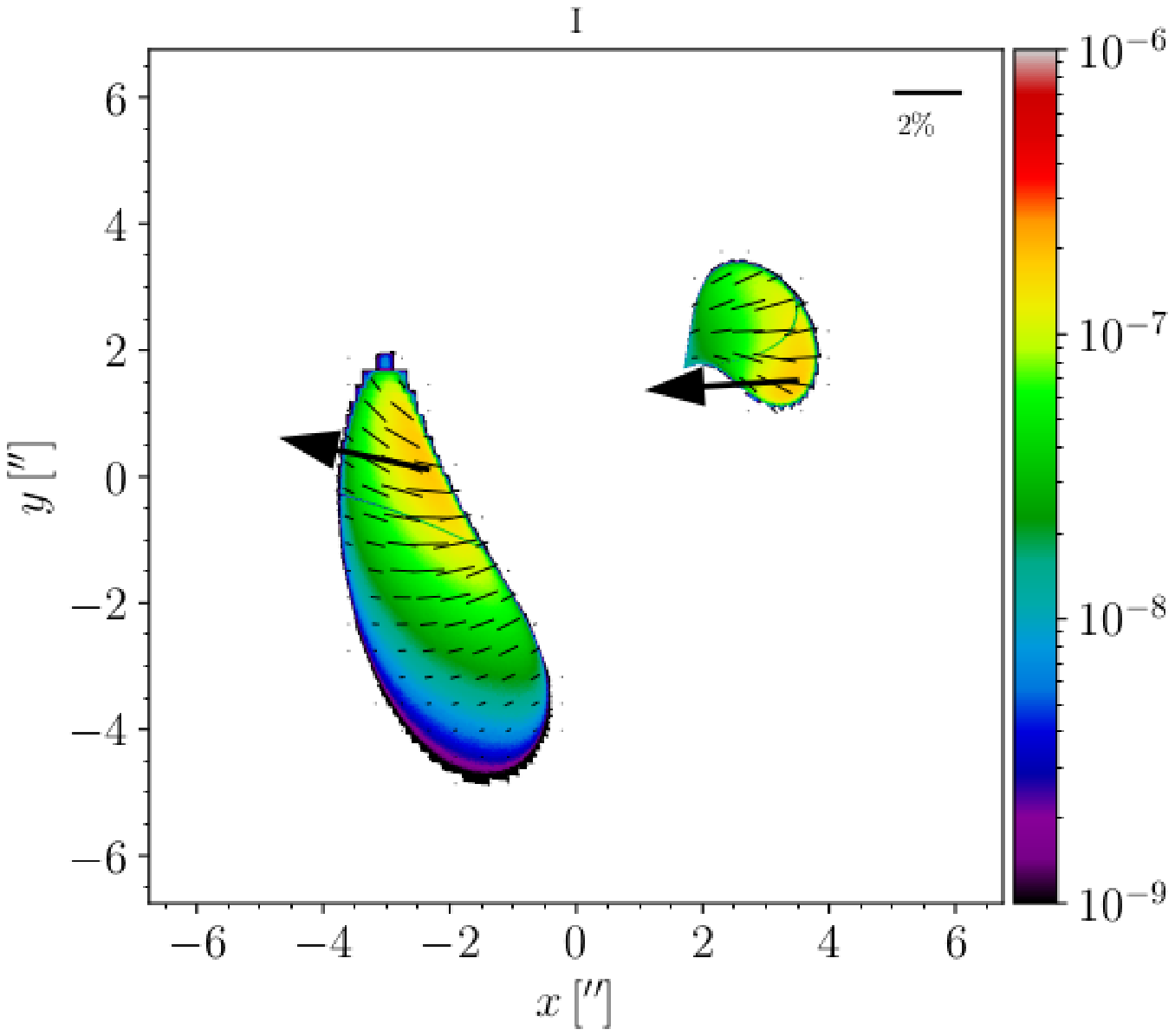}
\caption{The model of an extended jet-like polarized source (left) that is affected by the gravitational lens (right). The parameters for macro-lens are taken as for  Q0951+561 lens system. Black small arrows show the distribution and orientation of the electric vector (length corresponds to the rate of polarization as it is shown on the plots). The big arrows on the plots show only the orientation of the polarization vector, while the color presents intensity in arbitrary units (blue showing smaller intensities). }
\label{ml02}
\end{figure*}

The polarization in the continuum seems to be wavelength dependent, showing larger polarization degree at shorter wavelengths. The degree of polarization and polarization angle are different in A and B components. Previously, \cite{bo90} found that the polarization in both components should be 
$P \leq$ 3.2\% (taking 2$\sigma$ upper limit), that is  comparable with  our measurements which show the polarization  on the level of 1\%.

 {The observed continuum polarization in Q0957+561 can be due to  electron scattering  originating from the atmosphere of a  plane‐parallel scattering‐dominated disc.  In this case the  vector of electric field  is perpendicular to the symmetry axis of the disc, and the disc axis  is assumed to  be along the jet direction \citep[][]{ki03} 
However}, the polarization in the continuum can partly come from the central accretion disc, and partly from the synchrotron radiation of the optical continuum in the jet.

 If the polarized continuum is coming from the accretion disc, one can expect that the electric vector will be perpendicular to the jet.  Fig. \ref{fig-pol} shows that the vector of the electric field seems to be perpendicular to the projected jet direction of A  component, indicating that the polarization is probably originating in the accretion disc \citep[see e.g.][]{ki03}. However, the problem is to explain the wavelength dependent polarization if there is only polarization connected with the accretion disc. There are several ideas about the observed wavelength dependent polarization in some AGNs \citep[see][]{we93,be98}. {One possibility is  that polarization in the accretion disc, or in the 'hot corona'  (assumed to be around the disc) in the combination  with outflow  can give wavelength dependent polarization \citep[see][]{be98}. The Ly$\alpha$ of Q0957+561A,B shows P-Cyg profile \citep[see e.g.][]{do95,pop05} that indicates an outflow in hot gas,  supporting this scenario.}

\subsubsection{Polarization and gravitational lensing effect}

{ As we discussed in \S 3.2}, it is obvious that there are differences between the  polarization parameters of  A and B images. There is a difference in the PA of A and B components for around $\Delta\varphi_{AB} \sim 20^\circ$.  {The difference between the radio jet projections between these two components is smaller $\Delta\theta\sim 10^\circ $, but still exists \citep[see Fig. 5 in][]{go88}. This difference in the A and B jet angle projections can be seen also in \citet[][see their Fig 1]{ba99} and in \citet[][see their Fig. 2]{ha08}. } This may indicate two scenarios. The first is that the difference {   in polarization angle} is caused by macro-lens (and/or micro-lensing) effect of the continuum source. Otherwise, a difference of  polarization angle between the components may be caused by  the jet-disc precession, since the time delay between components A and B is around 420 days \citep[see][]{sha08,sha12}. The time delay may be  long enough to see two positions of the disc-jet system.

 We cannot expect that the gravitational lens  produces an additional polarization effect of a polarized source \citep[see][]{kr91}.
However, the gravitational distortion can change the observed polarization parameters, especially in the source where the polarization is depending from the dimensions of the emitting region 
\citep[as e.g. in the case of microlensing, see][]{pa20}. {In the radio-loud  quasars, the continuum polarization can have two contributing components, one that is coming from a disc and additionally a synchrotron polarization component that is coming from the jet. Considering the dimension of these two regions,  microlensing can affect the polarization component that is coming from disc \citep[the disc dimension is comparable with the Einstein radius ring - ERR for microlensing, see][]{pa20}  and macro-lensing may affect the polarization parameters coming from more extensive sources as e.g. the jet emission \citep[the jet dimension  can be comparable to ERR for macro-lensing, see][]{kr91}.  The quasar jet emission is the most intensive in the radio, but a smaller fraction of the jet (usually highly  polarized) emission can contribute to the optical part. }
Therefore one can expect that both gravitational macro- and micro-lensing may produce an additional effect in observed polarization of lensed images. 

In \citet{pa20} we demonstrate micro-lensing effects of polarized light due to the equatorial scattering, that qualitatively can be applied to other polarization mechanisms where the polarization parameters are depending on dimensions of polarization region or anisotropy of the polarization source.  Additionally, here we explore the influence of macro-lensing on the observed polarization in different images.

To demonstrate the influence of the strong lensing on the polarization signal, particularly on the polarization angle, we modeled the emitting region as an ellipsoidal source (jet-like structure, see Fig.  \ref{ml02} -  panels left). We assumed that the polarization vector has an orientation of $\alpha=90^\circ$  to the ellipsoid orientation. The polarization of the source follows the surface intensity (taken to have a distribution, the red color in Fig.  \ref{ml02} shows more intensive, and blue less intensive parts) and total polarization degree is $P=0.58$\%. The lens is represented with singular isothermal elliptic (SIE) potential \citep[for analytic forms for the SIE potential, see][]{ka93,ko94,ke98}.

  In Fig.  \ref{ml02} (right panel) we obtained two images of extended structure, with total polarization around     $P=1.42$\%  in both images, but the polarization angle between two images has a difference of $\Delta\alpha\sim 14^\circ$. Therefore, difference between the polarization angle in images is probably caused by macro-lens, but it is expected that polarization degree is similar in both images.

Note here that  the averaged polarization angles of components (shown as large arrows) do not  change significantly the orientation with respect to the polarization angle of the source, and therefore the observed polarization angle in both images is probably perpendicular to the observed radio jet. 
 { Here we used a simple model for  an extensive source in order to see distribution of the electric vector and find a total polarization angle, but similar effect can be expected if the polarization is combined from two source: one from a disc (which is dominant) and another highly polarized light from jet with a small contribution. As a final effect the polarization of the disc will be affected by the polarized light from the extensive jet which is differently amplified by macro-lensing. This may  be an explanation of observed    difference in polarization parameters of Q0957+561 A,B  images.}
  
   {Additionally, we cannot exclude that the  microlensing is causing the difference between polarization parameters in images A and B. The microlensing is affecting polarization of compact regions with anisotropic polarization. This effect is qualitatively similar to  the case of equatorial scattering shown in \citet{pa20}. However, it is  hard to disentangle macro-lensing and micro-lensing effects from one epoch (or two very near epochs) observation. Therefore, future polarization observations are needed to clarify this issue. }
  
  As we noted above, the second scenario   may be that the observed difference in the polarization angle is caused by the disc-jet precession.  {However, the VLBI observations obtained at different epochs show the same orientation of the jet in images A and B \citep[see][]{ha97,ha08}, therefore it is  unlikely that the  jet  precession is present.}

\section{Conclusions}

Here we presented spectroscopic and polarimetric observations of the lensed quasar Q0957+561 obtained with the 6m SAO RAS telescope. We analyzed our observations from two epochs, and we compare our observations with previous ones. From our analysis we can conclude the following:

\begin{itemize}
    \item The B/A ratio during both epochs was around two, which indicates a strong magnification of the component B. Both images show bluer spectrum as brightness becomes stronger, this effect is probably mostly caused by the intrinsic variability in the quasar. However, there is a difference in this change in the component B compared to the component A. { The interval in the change of the component B seems to be larger than in the component A}. This indicates that, in addition to the intrinsic variation, microlensing probably contributes to brightness of the component B.
    \item Polarization in the broad lines is not present, and it is (within the  error-bars) on the level of the continuum. Therefore, the equatorial scattering probably is not dominant in the broad lines. It may indicate two effects: the first is  the complete lack of equatorial scattering, and the second is the presence of a depolarization region above the BLR. Moreover, an absorption observed in the Ly$\alpha$ line indicates an outflowing BLR, which may be a depolarization region located between the observer and BLR.  {An alternative scenario for the lack of polarization in the broad lines is that the inner equatorial scattering region is  comparable with an outer BLR radius.}
    \item The polarization in both components seems to be wavelength dependent, and polarization vector { is almost} perpendicular to the observed radio-jet. This indicates that the continuum  polarization may come from the accretion disc, and that there are some effects which are causing the wavelength dependent polarization  \citep[see e.g.][]{we93,be98}. 
    \item The polarization parameters between  Q0957+561 A and B components are  different. Using a sample model of polarized extensive source,  we show that gravitational { macro-}lensing could explain these differences.  {However, we cannot exclude some other effects, as e.g. microlensing.}
    \item Using Mg II FWHM and $\lambda L(3000\AA)$  from the high quality spectrum of the component A, we obtained that the Q0957+561 SMBH mass is $M_{\rm SMBH}\approx (4.8-6.1) \cdot 10^8M\odot$ 
\end{itemize}

The polarization effect in lensed quasar Q957+561 seems to have different nature than in SDSS J1004+411, and it seems that in the case of Q957+561   the macro-lensing  effect contributes to the detected difference in the polarization angle between component A and B. 

\section{Acknowledgments}

This work is supported by the Ministry of Education, Science and Technological Development of R. Serbia  (the contract 451-03-68/2020-14/200002). 
VLA and ESS thank the grant of Russian Science
Foundation project number 20-12-00030 ‘Investigation of geometry
and kinematics of ionized gas in active galactic nuclei by polarimetry
methods’, which supported the spectropolarimetric and polarimetric observations and data analyze.
Observations with the SAO RAS telescopes are supported by the Ministry of Science and Higher Education of the Russian Federation (including agreement No05.619.21.0016, project ID RFMEFI61919X0016).  {We would like to thank the referee  for giving very useful comments which  helped improving the quality of the paper.}

%
%

\end{document}